\documentclass[aps,pre,twocolumn,groupedaddress,superscriptaddress,showpacs]{revtex4-1}

\usepackage{amsmath}
\usepackage{graphicx}
\usepackage{float}
\usepackage{color}

\begin{document}

  \title{Kinetic roughening of aggregates of patchy colloids with strong and weak bonds}

  \author{C. S. Dias}
   \email{csdias@fc.ul.pt}
    \affiliation{Departamento de F\'{\i}sica, Faculdade de Ci\^{e}ncias, Universidade de Lisboa, P-1749-016 Lisboa, Portugal, and Centro de F\'isica Te\'orica e Computacional, Universidade de Lisboa, Avenida Professor Gama Pinto 2, P-1649-003 Lisboa, Portugal}

  \author{N. A. M. Ara\'ujo}
   \email{nmaraujo@fc.ul.pt}
   \affiliation{Departamento de F\'{\i}sica, Faculdade de Ci\^{e}ncias, Universidade de Lisboa, P-1749-016 Lisboa, Portugal, and Centro de F\'isica Te\'orica e Computacional, Universidade de Lisboa, Avenida Professor Gama Pinto 2, P-1649-003 Lisboa, Portugal}

  \author{M. M. Telo da Gama}
   \email{margarid@cii.fc.ul.pt}
    \affiliation{Departamento de F\'{\i}sica, Faculdade de Ci\^{e}ncias, Universidade de Lisboa, P-1749-016 Lisboa, Portugal, and Centro de F\'isica Te\'orica e Computacional, Universidade de Lisboa, Avenida Professor Gama Pinto 2, P-1649-003 Lisboa, Portugal}


\begin{abstract}
We study the irreversible aggregation of films of patchy spherical colloids with directional and selective
interactions. We report a crossover of the interfacial roughening 
from the \textit{Kardar-Parisi-Zhang} (KPZ) to the KPZ with quenched disorder (KPZQ) universality class 
when the difference between the strong and weak bonds is sufficiently large. We calculate the critical exponents and 
identify the crossover between the two regimes.
\end{abstract}
 
  \maketitle

\section{Introduction}

A very active field, in soft matter, is the experimental and theoretical study of the nonequilibrium scaling of
growing interfaces 
\cite{Takeuchi2014,Takeuchi2011,Takeuchi2012,Takeuchi2013,Takeuchi2010,Wakita1997,Hallatschek2007,Huergo2010,Huergo2011,Huergo2012,Santalla2014,Alava2004,Sakaguchi2010,Vergeles1995,Kim2001}.
In particular, recent work on the aggregation of colloids at the edge of an evaporating drop, revealed that the interfacial roughening
depends on the asymmetry of the colloidal shape. The 
experimental results suggest a Poisson-like growth for spherical colloids while for strongly anisotropic 
colloids the growth is in the universality class of \textit{Kardar-Parisi-Zhang} with quenched disorder (KPZQ).
For moderate anisotropy the classical \textit{Kardar-Parisi-Zhang} (KPZ) universality class is observed \cite{Yunker2013,Yunker2013b,Nicoli2013}. Here, we 
consider the asymmetry of spherical colloids by including regions on the colloidal surface with anisotropic
interactions (patches).

\begin{figure}[t]
   \begin{center}
   \includegraphics[width=\columnwidth]{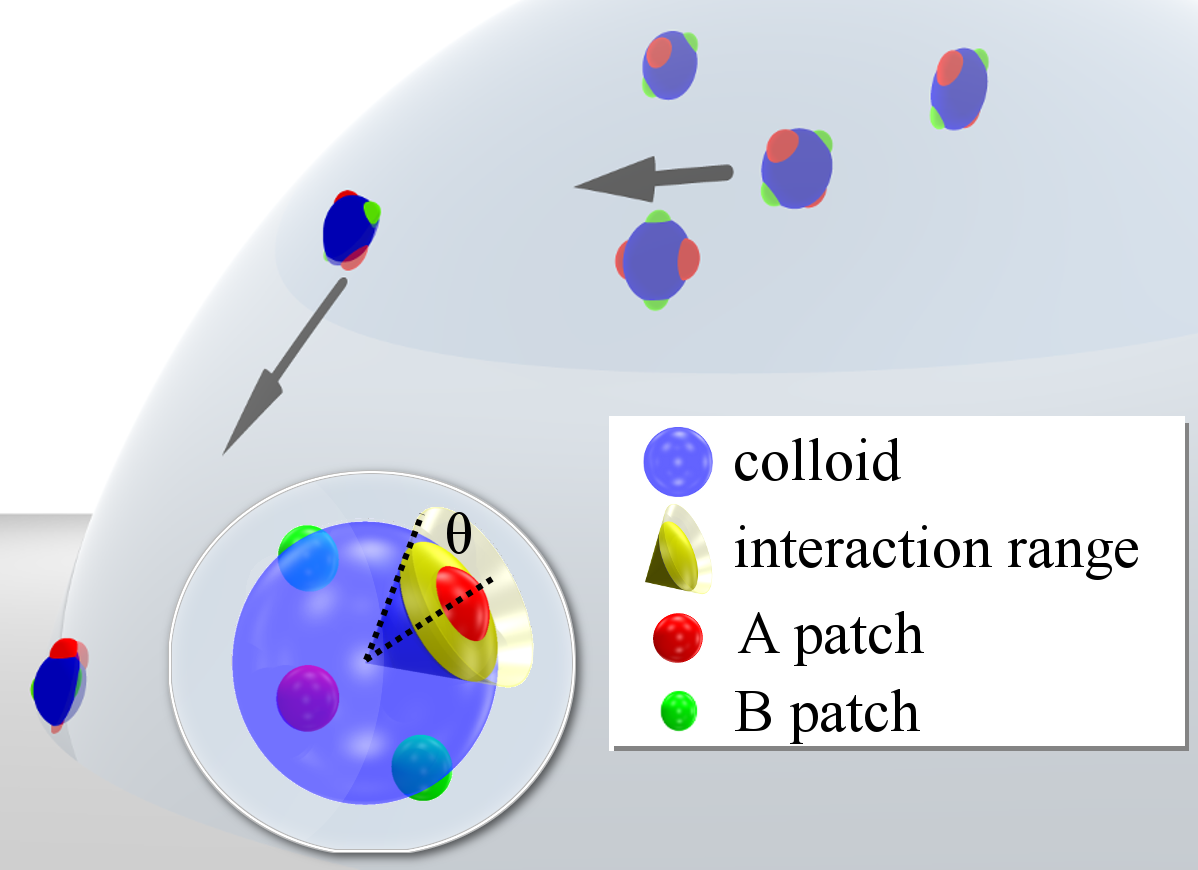} \\ 
   \caption{Illustration of the aggregation at the edge of an evaporating drop of patchy colloids with distinct bonding sites. The flow due to evaporation
drags the colloids towards the edge where they either adsorb on the substrate or bind to another colloid. 
Magnified colloid: Schematic representation of a patchy colloid (blue) with two A-patches (red) and two B-patches (green) on its surface and the respective interaction range (yellow). 
The interaction range is truncated at an angle $\theta$ with the center of the patch.}
    \label{fig.drop}
   \end{center}
   \end{figure}

The \textit{Kardar-Parisi-Zhang} (KPZ) is a very robust universality that describes successfully many growing interfaces \cite{Kardar1986}.
Underlying this class is the stochastic differential equation,

 \begin{equation}
     \frac{\partial h(x,t)}{\partial t}=\nu\frac{\partial^2 h(x,t)}{\partial x^2}+\lambda\left(\frac{\partial h(x,t)}{\partial x}\right)^2+\eta(x,t), \\
     \label{eq:kpz}
 \end{equation} 
where $h(x,t)$ is the height of the interface at the lateral position $x$ and time $t$. The first term on the right-hand side is the interface relaxation 
due to surface tension $\nu$ and the non-linear term accounts for lateral growth \cite{Barabasi1995}. The noise term $\eta(x,t)$ has zero average. 
If the noise does not depend on time (quenched noise) but on the lateral position and height, $\eta(x,h)$, 
the growth falls into another universality class, the KPZ with quenched disorder (KPZQ) \cite{Csahok1993,Leschhorn1996,Barabasi1995}. 
This class can be described by the differential equation,

 \begin{equation}
     \frac{\partial h(x,t)}{\partial t}=\nu\frac{\partial^2 h(x,t)}{\partial x^2}+\lambda\left(\frac{\partial h(x,t)}{\partial x}\right)^2+F+\eta(x,h). \\
     \label{eq:kpzq}
 \end{equation} 
Here a constant driving force $F$ must be included to keep the interface from pinning when $\eta(x,h)<0$.
 
The last two decades have been fruitful on models and experimental setups developed to shed light into the
KPZ universality with quenched disorder \cite{Olami1994,Csahok1993,Leschhorn1997,Amaral1995,Amaral1994,Buldyrev1992,Halpin-Healy1995,Tang1992}.
From these studies, two main results emerged: First, below a critical driving force, the interface is pinned and above it 
depinned. Second, the scaling of the interface above the critical driving force, is initially in the KPZQ class but it does 
crossover to the KPZ class for sufficiently large length and/or long time scales. Solely at the critical driving force is KPZQ observed, in the thermodynamic limit.

The new ingredient in our study is the use of colloids with anisotropic regions distributed on their surface (patches) yielding
highly directional pairwise interactions. This type of colloids, called patchy colloids, can be synthesized 
using many different techniques \cite{Yi2013,Wilner2012,Hu2012,Duguet2011,Pawar2008,Shum2010,Wang2012,He2012}. 
Along the experimental advances, significant progress has been made theoretically. 
However, theoretical understanding has been focused mostly on the study of equilibrium 
phase diagrams \cite{Glotzer2004, Ruzicka2011,Bianchi2006,Russo2010}. More recently, the kinetics of aggregation \cite{Vasilyev2013,Vasilyev2014} and self-organization on substrates 
was addressed \cite{Bernardino2012,Dias2013,Dias2013b,Dias2013a,Gnan2012,Dias2014}, and the nonequilibrium adsorption revealed interesting
behavior for single type colloids \cite{Dias2013} mixtures of colloids \cite{Dias2013b}, or colloids with selective interactions \cite{Dias2013a}.

Patchy colloids with distinct patch-patch interactions yield interesting properties in the bulk, such as 
a vanishing critical point \cite{Tavares2009,Tavares2009a} as well as unusual thermodynamic and percolation 
properties \cite{Tavares2010a,Tavares2010}. Usually, $2AnB$ colloids are considered, with strong A- and weak B-patches
(two $A$-patches in the poles and $n$ $B$-patches along the equator). When two A-patches form sufficiently strong bonds, compared to AB patches (BB patches do not interact),
the equilibrium liquid-vapor binodal was found 
to be re-entrant both in continuum \cite{Russo2011,Russo2011a} and lattice models in two and three dimensions \cite{Almarza2011, Almarza2012,Tavares2014}. 
Here, we consider $2A2B$ colloids where the $\mathrm{AA}$, $\mathrm{AB}$, and $\mathrm{BB}$ 
binding probabilities are dissimilar, and BB patches do interact. By contrast to what was observed experimentally for spherical colloids (without patches) \cite{Yunker2013}, 
Poisson-like growth is never observed. Depending on the relation between binding probabilities the interface is either in the KPZ or KPZQ universality 
class with a crossover between the two.

In this work we bring together the fields of interfacial growth and functionalized colloids to address the effect of 
patch-like anisotropy with short-ranged interactions on the nonequilibrium scaling of a growing interface.

\section{Model}\label{sec:model}

In an evaporating drop, colloids are dragged to its edge 
(see fig.~\ref{fig.drop}), and thus we consider advective transport of colloids towards the substrate. 
Numerically, we consider a bidimensional system where the contact between the 
edge of the drop and the substrate is a straight line. 
Similar to the transport mechanism used in Ref.~\cite{Yunker2013}, 
colloids are released far away from the substrate at a random lateral 
position and a vertical straight trajectory towards the substrate is considered.
Colloids irreversibly adsorb at the substrate or bind to other colloids.
Interaction with the substrate is isotropic and colloids bind to it with a random orientation.

\begin{figure*}[t]
   \begin{center}
   \includegraphics[width=1.8\columnwidth]{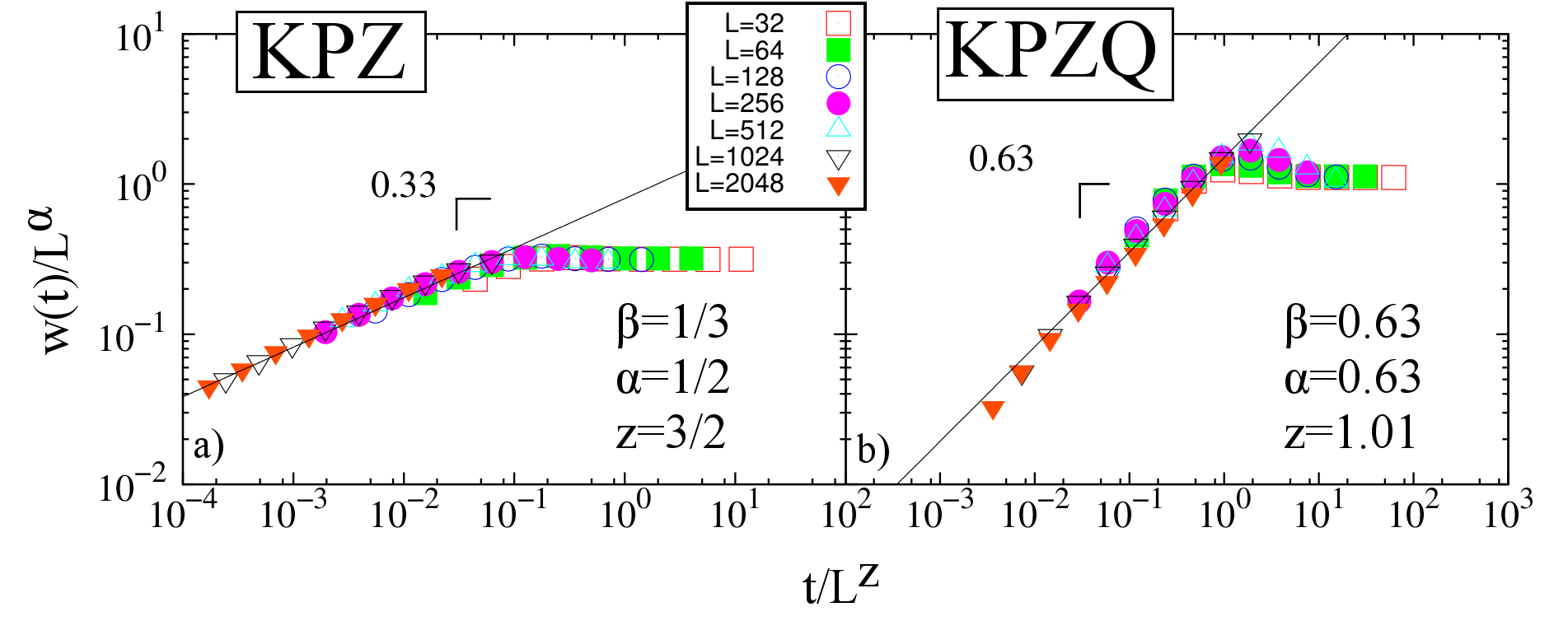} \\ 
   \caption{Roughness as a function of time for the aggregation of $2A2B$ colloids at the edge of a drop.
   Data collapse according to the Family-Vicsek scaling relation (\ref{eq:family}) for (a) $r_\mathrm{AB}=1$ and 
   (b) $r_\mathrm{AB}=0.01$. Simulations were performed on substrates of linear size ranging from $L=32$ to $L=2048$ with $2048L$
   patchy colloids averaged over $320000$ samples, for the smaller systems, and $20000$ samples for the larger ones.}
    \label{fig.rough_KPZQ}
   \end{center}
   \end{figure*}
   
To describe the colloid/colloid interaction we use the model introduced in Ref.~\cite{Dias2013}, where an interaction range is defined as a region 
around the patch parameterized by an angle $\theta=\pi/6$ with the center of the patch (see magnified colloid in fig.~\ref{fig.drop}). In the event of a 
collision with another colloid, the binding occurs only if both interaction ranges overlap. For a successful binding, the probability is
checked taking into account the binding probability between the specific patches, and, if binding occurs, the newly aggregated 
colloid reorients itself so that both patches are aligned. 
We consider selective interaction between patches, described by distinct binding probabilities.

\begin{figure}[t]
   \begin{center}
    \includegraphics[width=\columnwidth]{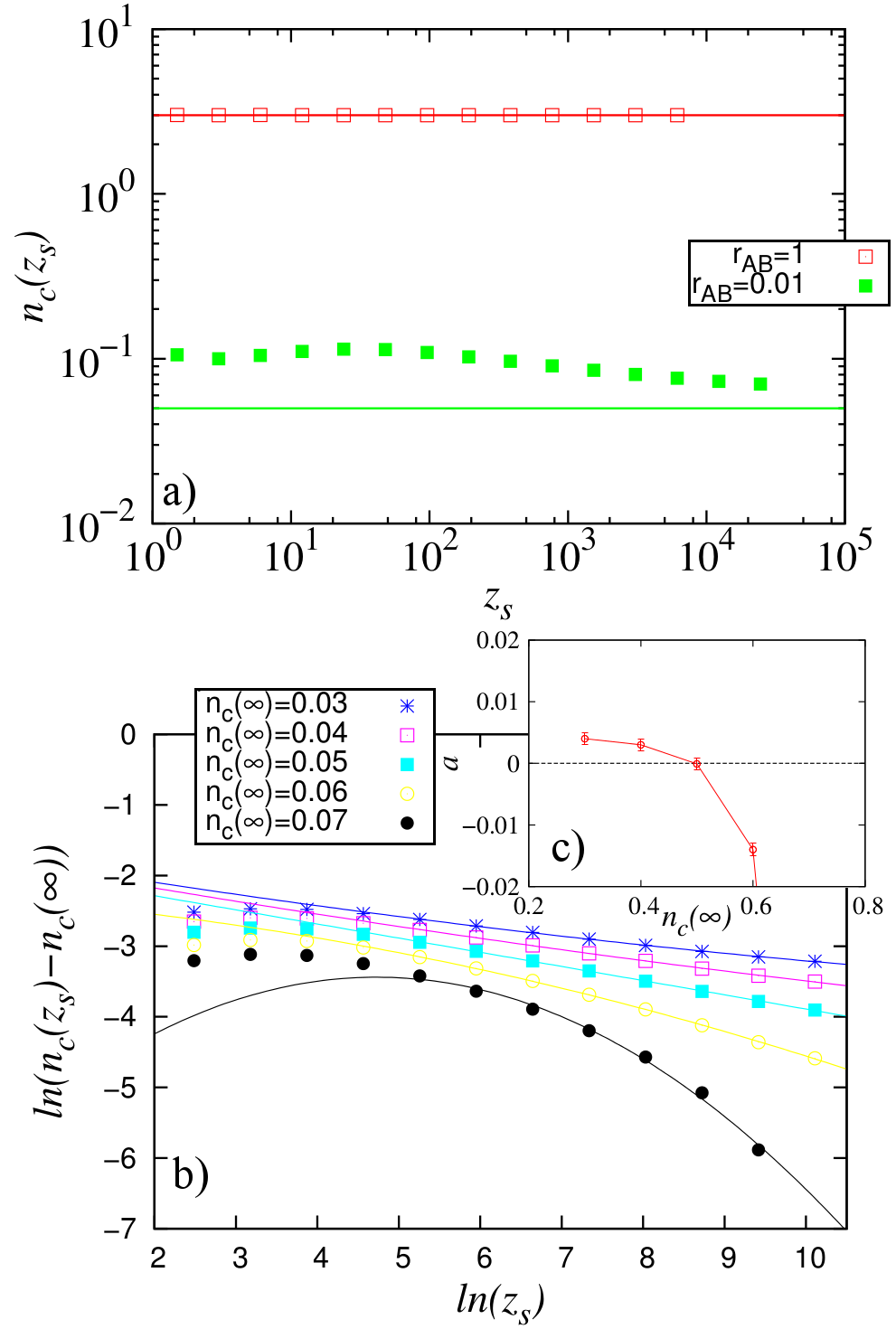} \\
\caption{Effect of $r_\mathrm{AB}$ on the number of bonds $\mathrm{AA}$, $\mathrm{AB}$, and $\mathrm{BB}$. 
(a) Ratio of weak to strong bonds $n_c=(N_\mathrm{AB}+N_\mathrm{BB})/N_\mathrm{AA}$ as a function of
the distance to the substrate $z_s$ for $r_\mathrm{AB}=\{0.01,1\}$ and a system of size $L=4096$ and $8192L$ colloids. (b) Scaling
of $n_c(z_s)-n_c(\infty)$ for values of $n_c(\infty)=\{0.03,0.05,0.06,0.07\}$ at $r_\mathrm{AB}=0.01$ and a system of size $L=4096$ and $8192L$ colloids.
The fitting is polynomial of the type $a\ln(z_s)^2+b\ln(z_s)+c$ on a log scale . When $a=0$, $n_c(z_s)$ is fitted by a power law. As shown 
in the inset (c) this is observed for $n_c(\infty)=0.05\pm0.01$.}
  \label{fig.connections}
   \end{center}
  \end{figure}

We assume chemical bonding between the patches, \textit{i.e.} highly directional and irreversible within the timescale of interest \cite{Leunissen2011}.
We also assume that the process of establishing a chemical bond is thermally activated and characterized by an activation barrier $E_a$.
In general, this barrier is different for 
$\mathrm{AA}$, $\mathrm{AB}$, and $\mathrm{BB}$ bonds. For simplicity, we assume that the higher activation barrier for $\mathrm{BB}$ is the same as for $\mathrm{AB}$ bonds. The rate of the binding process 
$i=\{\mathrm{AA}, \mathrm{AB}, \mathrm{BB}\}$ is Arrhenius-like,

    \begin{equation}
      P_i\propto e^{-E_a^{i}/{k_BT}},\\
      \label{eq:arrhenius}
    \end{equation} 
where $k_B$ is the Boltzmann constant and $T$ is the thermostat temperature. We assume that the prefactor depends only on the frequency of patch-patch collisions 
and is independent of the patch types. The binding probability 
of $\mathrm{AA}$, $\mathrm{AB}$, and $\mathrm{BB}$ bonds upon colliding is then $P_\mathrm{AA}$, $P_\mathrm{AB}$, and $P_\mathrm{BB}$. 
Without loss of generality we consider $P_\mathrm{AA}=1$ and for simplicity $P_\mathrm{AB}=P_\mathrm{BB}$,
and define the sticking coefficient $r_\mathrm{AB}=P_\mathrm{AB}/P_\mathrm{AA}$. $\mathrm{AA}$ bonds are favored for 
low $r_\mathrm{AB}$ that promotes the growth of chains.

\section{Results}\label{sec:results}

We performed simulations for substrates of linear sizes ranging from $L=32$ to $L=4096$ in units of the colloid diameter $\sigma$ and
adsorption and binding of as many as $8192L$ colloids.

To study the nonequilibrium scaling of the growing interface we need to compute its morphology. To characterize 
this morphology we calculate the interfacial roughness $w$. We divide the system in $N$ vertical
columns of width $\sigma$, where $N=L/\sigma$. For each column $i$ we simulate a downward trajectory of a colloid 
released from above the maximum height of the film and calculate the height $h_i$ where it
collides with either one colloid or the substrate. The
roughness at time $t$ is then defined as,

\begin{equation}\label{eq.roughness}
w(t)=\sqrt{\langle\left[h_i(t)-\langle h(t)\rangle\right]^2\rangle}, \\
\end{equation}
where $\langle h(t)\rangle=\sum_i{h_i}/N$ is averaged over the $N$ columns. Here, the time $t$ is defined as the number of absorbed layers of colloids (equivalent 
to the experimentally used average height). After the saturation time $t_\mathrm{sat}$, the correlation length perpendicular to the growth direction
is of the order of the system size and the roughness saturates at
$w=w_\mathrm{sat}$ \cite{Odor2004,Barabasi1995}. Both the saturation roughness and saturation time scale with the system size as 
$w_\mathrm{sat}\sim L^{\alpha}$ and $t_\mathrm{sat}\sim L^z$, where
$\alpha$ is the roughness exponent and $z$ is the dynamical exponent. The short-time behavior of the interface roughness is also a power law given by
$w(t)\sim t^\beta$ where $\beta$ is the growth exponent. The interface can then be described by the \textit{Family-Vicsek} \cite{Family1985} scaling relation,

  \begin{equation}
     w(L,t)=L^\alpha f\left(\frac{t}{L^z}\right), \\
     \label{eq:family}
  \end{equation} 
  where $f(u)$ is a scaling function. Using the scaling relation and the exponents for different universality classes we can identify 
  the universality class of the growing interface.  
   
In fig.~\ref{fig.rough_KPZQ} the data collapse of the rescaled 
roughness is shown, for two limiting values of the sticking coefficient $r_\mathrm{AB}=\{0.01,1\}$. For systems with $r_\mathrm{AB}=1$, 
fig.~\ref{fig.rough_KPZQ}a), data collapse is obtained using the critical exponents of the KPZ universality class, 
namely $\beta=1/3$, $\alpha=1/2$, and $z=3/2$. However, for 
$r_\mathrm{AB}=0.01$, fig.~\ref{fig.rough_KPZQ}b), data collapse is only observed for the critical exponents of the KPZQ 
universality class $\beta=0.63$, $\alpha=0.63$, and $z=1.01$. For other values of $r_\mathrm{AB}$ we have found that the colloidal network interface 
is either in the KPZ or KPZQ universality class, for large and small sticking coefficients, respectively.

For $P_\mathrm{AA}>P_\mathrm{AB}=P_\mathrm{BB}$, due to the presence of two strong bonding sites, growth is promoted along the poles, 
favoring the aggregation of AA chains. These chains are not necessarily aligned vertically and may extend over long lateral regions, 
blocking the access of new colloids to the underlying colloids. These regions will only 
have B-patches available for binding, and thus the probability of binding (growth) there is lower. 
This is expected to have an effect similar to that of quenched noise in eq.~(\ref{eq:kpzq}).

This type of quenched noise also affects the film structure,
as shown in fig.~\ref{fig.connections}a), for the ratio between weak bonds $\mathrm{AB}$/$\mathrm{BB}$ and 
strong bonds $\mathrm{AA}$, $n_c=(N_\mathrm{AB}+N_\mathrm{BB})/N_\mathrm{AA}$ as a function of
the distance $z_s$ to the substrate. We note that for $r_\mathrm{AB}=0.01$ the effect of the substrate is 
observed for the entire range of $z_s$, as evident from the decrease of $n_c$ with $z_s$. By contrast, in systems with $r_\mathrm{AB}=1$, 
$n_c$ saturates at a non-zero value. For systems with low $r_\mathrm{AB}$, and interfacial roughening in the KPZQ class, in fig.~\ref{fig.connections}b), 
we found that the ``liquid film'' never saturates. Rather, we observe a power-law 
scaling of $n_c(z_s)-n_c(\infty)$ as a function of $z_s$, where $n_c(\infty)$ is the ratio of bonds in the limit of infinite thickness,
and $n_c(\infty)=0.05\pm0.01$ for systems with $r_\mathrm{AB}=0.01$. To calculate $n_c(\infty)$, a polynomial fit of $\ln(n_c(z_s)-n_c(\infty))$ 
as a function of $\ln(z_s)$ of the 
form $a\ln(z_s)^2+b\ln(z_s)+c$ was used. A linear dependence is recovered when $a=0$, giving $n_c(\infty)=0.05\pm0.01$, for $r_\mathrm{AB}=0.01$ (see fig.~\ref{fig.connections}c)).
   
 \begin{figure}[t]
    \begin{center}
    \includegraphics[width=\columnwidth]{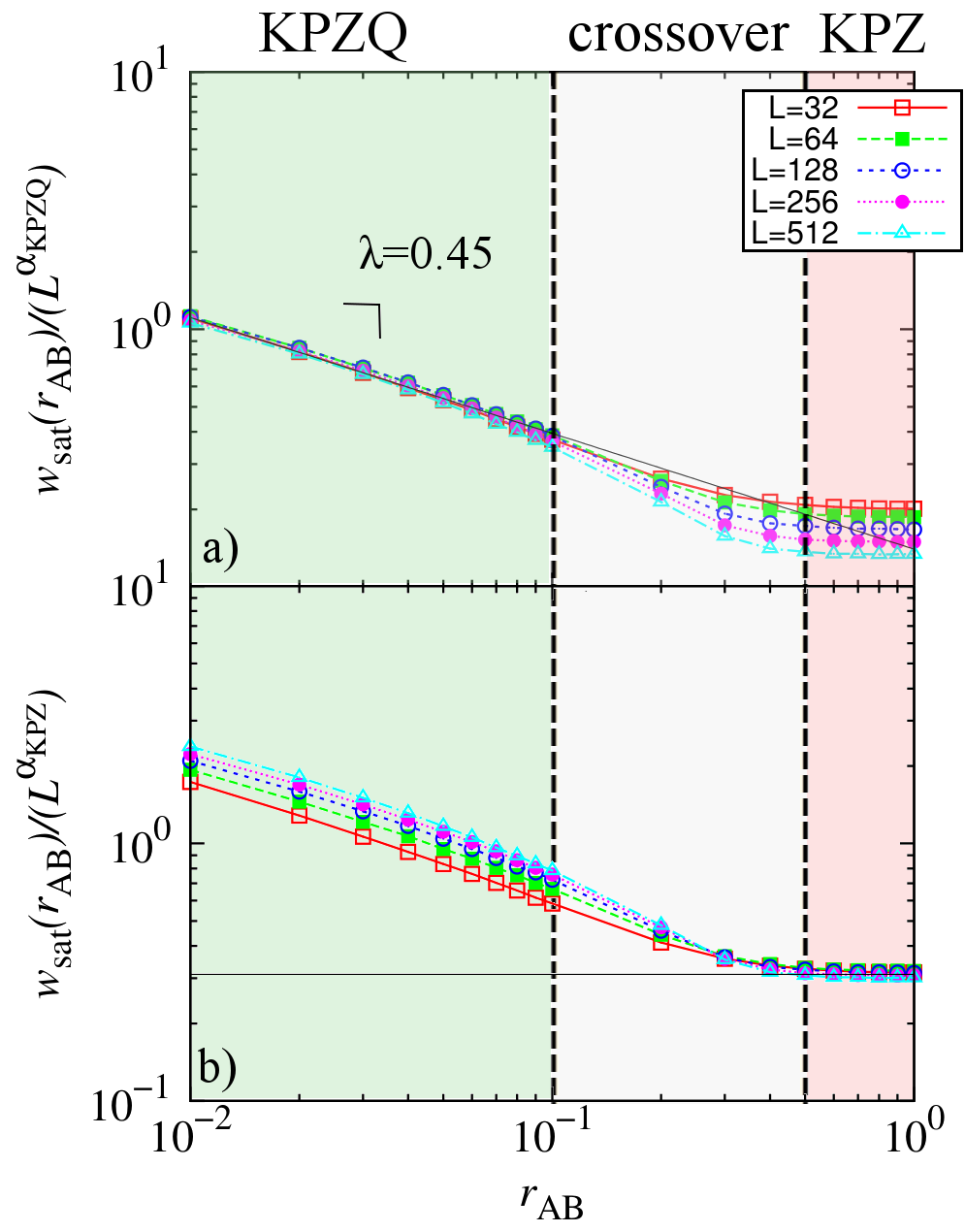} \\ 
    \caption{Transition from KPZ to KPZQ universality. 
    Rescaled saturation roughness as a function of the sticking coefficient $r_\mathrm{AB}$. Data collapse for system sizes 
    ranging from $L=32$ to $L=512$ of $w_\mathrm{sat}\left(r_\mathrm{AB}\right)L^{-\alpha}\sim f\left(r_\mathrm{AB}\right)$ 
    for (a) the roughness exponent of KPZQ $\alpha=0.63$ and (b) the roughness exponent of KPZ $\alpha=0.5$. Results 
    averaged over $320000$ samples for the smaller systems and $80000$ samples for the larger ones.
    }
     \label{fig.rough_sat}
    \end{center}
    \end{figure}   
   
Next we investigate how the behavior of the interfacial growth crosses from KPZ to KPZQ by considering different values
of the sticking coefficient $r_\mathrm{AB}$. For low sticking coefficients, mainly $\mathrm{AA}$ bonds are formed
and the growth is dominated by chains. 
Increasing the sticking coefficient, the number of junctions increases, the effect of lateral growth becomes dominant, and the KPZ universality class is recovered.

 \begin{figure}[t]
    \begin{center}
    \includegraphics[width=\columnwidth]{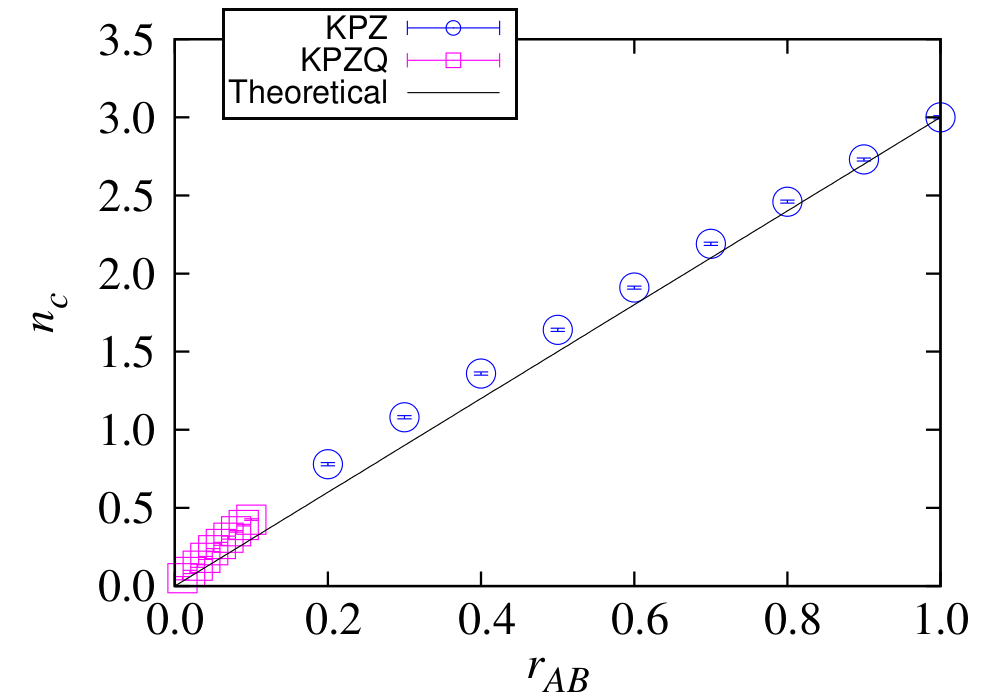} \\ 
    \caption{Ratio of bonds in the thermodynamic limit $n_c(\infty)$ as a function of $r_\mathrm{AB}$. The black solid line corresponds 
    to the mean-field relation $n_c(\infty)=3r_\mathrm{AB}$.
    }
     \label{fig.plateau}
    \end{center}
    \end{figure}   

Figure~\ref{fig.rough_sat} shows the saturation roughness $w_\mathrm{sat}$ rescaled by $L^\alpha$ as a function
of the sticking coefficient $r_\mathrm{AB}$. In fig.~\ref{fig.rough_sat}a), the rescaling is done using the roughness exponent of the KPZQ
universality class, $\alpha=0.63$. It is clear that at low values of $r_\mathrm{AB}$ the curves collapse indicating that the systems 
belong to this universality class. Note in fig. \ref{fig.rough_KPZQ}b) that the 
data collapse is observed over two orders of magnitude in size making the possibility of a finite-size crossover very unlikely. 
A power-law dependence of the saturation roughness on the sticking coefficient, $w_\mathrm{sat}\sim r_\mathrm{AB}^{-\lambda}$ is 
observed, in the KPZQ region, with an exponent 
$\lambda=0.45\pm0.01$. This decrease is related to an increase of the number of junctions which
promotes lateral growth and consequently hampers roughening. For systems with $r_\mathrm{AB}>0.1$ no data collapse is observed with the KPZQ exponents. 

Figure~\ref{fig.rough_sat}b), reveals that using the roughness exponent of KPZ, $\alpha=1/2$, 
data collapse is observed for $r_\mathrm{AB}>0.5$, indicating that above that threshold the interface falls into the KPZ universality class. 
By contrast with the KPZQ systems, the saturation roughness $w_\mathrm{sat}$ does not depend significantly on the sticking
coefficient $r_\mathrm{AB}$. 

The fact that KPZQ is observed over an extended region of $r_\mathrm{AB}$ is remarkable since, for previous models of interfacial growth, KPZQ is only 
found at a critical value of the control parameter (\textit{e.g.}, the force $F$ in eq.~(\ref{eq:kpzq})). The reason why the interface is critical for 
different values of $r_\mathrm{AB}$ is likely related to two competing mechanisms that counterbalance to keep the system at criticality. 

The lower the value of $r_\mathrm{AB}$ the lower is the probability of 
binding to a B-patch. However, this also promotes the growth of long AA-chains and these long chains 
have more available B-patches, compensating the decrease in the binding probability. Figure~\ref{fig.plateau} shows the dependence 
on $r_\mathrm{AB}$, of the ratio of weak to strong bonds in the limit of infinite thickness, $n_c(\infty)$. The black-solid line is 
the mean-field dependence, given by, $n_c(\infty)=3r_\mathrm{AB}$, where we used the fact 
that $P_\mathrm{AB}=P_\mathrm{BB}$. We have found that the ratio of bonds is always 
above the mean-field limit supporting the idea that longer AA chains promote the bonding of B sites, beyond what is expected in the absence of correlations. 
It is also noteworthy that the typical size of the chains is much smaller than the system size. For example, 
for systems with $r_\mathrm{AB}=0.01$ the mean-field value of $n_c(\infty)=0.03$, implies that the average size of the AA chains is of the order of thirty colloids.

\section{Conclusions}\label{sec:conclusions}

We present a model where we control the assembly of chains or junctions by means of the selective interaction between A and B patches. For 
similar interactions ($r_\mathrm{AB}\approx 1$), there is no preferential bonding and junctions are likely to form. For low $r_\mathrm{AB}$
($r_\mathrm{AB}\ll1$), $\mathrm{AA}$ bonds are favored and the growth is dominated by chains. For $r_\mathrm{AB}=0$ only chains grow and no 
interface is formed. One can see the latter limit as a pinned phase with the depinned phase corresponding to the region of KPZ, with a critical 
depinning transition for the entire region of KPZQ.

The prototypical example of interfacial growth and depinning transitions is the imbibition of a fluid through a randomly disordered porous medium 
\cite{Buldyrev1992,Leschhorn1997,Alava2004,Csahok1993a,Morais2011}. Typically, a fluid invades the pores sequentially 
as far as the force (fluid pressure) is above the threshold for each pore. While below the critical force invasion 
is suppressed at a finite distance to the entrance, above it invasion is perpetually sustained and the interface 
is in the KPZ universality class. Only at the critical force the KPZQ universality class is recovered. This universality class 
for the interface is then intimately related to a direct percolation transition in the bulk \cite{Alava2004}. In our case, 
the growth occurs due to random addition of colloids to the interface. In order for the interface to grow, colloids 
need to find their way to the available patches. Since there are two patches on the poles and two along the equator, growth of 
individual branches can only be suppressed by steric effects of other branches, 
but the interface as a whole will always grow \cite{Dias2014}. Thus, one does not expect any pinned phase for $r_\mathrm{AB}>0$. 
However, for $r_\mathrm{AB}\ll1$ growth is suppressed in regions with a sufficiently large fraction of B-patches and thus KPZQ is obtained. 
For $r_\mathrm{AB}=1$ patches are similar and the interface is in the KPZ universality class. KPZQ is only found at the depinning transition, 
which in general occurs at a well defined threshold. Here, we found that the critical depinning regime occurs over a finite range of $r_\mathrm{AB}$. 

Our results have two consequences of practical interest. First, the roughness of the interface 
can be effectively controlled by the sticking coefficient. 
Second, our study opens the possibility for an experimental realization of a system with 
an extended region in the KPZQ universality class, without fine tuning of the control parameters. Possible experimental realizations include the
aggregation of colloidal particles functionalized by DNA \cite{Geerts2010,Wang2012} at the edge of a drop \cite{Yunker2013,Yunker2013b} or 
in other two-dimensional geometries \cite{Iwashita2013,Iwashita2014}.

\begin{acknowledgments} 
We acknowledge financial support from the
Portuguese Foundation for Science and Technology (FCT) under Contracts
nos. EXCL/FIS-NAN/0083/2012, PEst-OE/FIS/UI0618/2014, and IF/00255/2013.
\end{acknowledgments}

\bibliography{paper}

\end{document}